\pdfoutput=1
\documentclass[11pt]{article}
\usepackage[margin=1in]{geometry}
\usepackage[T1]{fontenc}
\usepackage[utf8]{inputenc}
\usepackage{lmodern,microtype,amsmath,amssymb,booktabs,tabularx,graphicx,xcolor,float,longtable,array,pdflscape,authblk}
\usepackage[section]{placeins}
\usepackage{caption}
\usepackage[numbers,sort&compress]{natbib}
\usepackage[colorlinks=true,allcolors=blue]{hyperref}

\graphicspath{{figures/}}
\input{si_labels}   

\title{TCSP3.0: Template-Based Crystal Structure Prediction with Coefficient-Preserving Element Substitution and Structural Diversity}
\author[1]{Ying Feng}
\author[2]{Jeffrey Hu}
\author[3]{Lai Wei}
\author[1]{Liang Tao}
\author[1]{Madeline Miklos}
\author[1]{Dipannoy Das Gupta}
\author[1]{Hossein Erfanshekooh}
\author[1,*]{Jianjun Hu}
\affil[1]{Department of Computer Science and Engineering, University of South Carolina, Columbia, SC 29201, USA}
\affil[2]{Department of Materials Science and Engineering, University of Illinois Urbana-Champaign, IL, USA}
\affil[3]{Department of Computer Science and Cybersecurity, University of North Georgia, Dahlonega, GA 30597, USA}
\affil[*]{Corresponding author: \texttt{jianjunh@cse.sc.edu}}
\date{}

\begin{document}
\maketitle
\begin{abstract}
Template-based crystal structure prediction (CSP) transfers a known crystal structure to a new composition by element substitution, so its accuracy depends on retrieving suitable templates, mapping elements and ranking the resulting candidates correctly. Auditing the original TCSP2.0 implementation on the T180 benchmark, we found that 469 of its 2,347 selected structures (20\%) have the wrong composition, caused by element mappings that ignore stoichiometric coefficients, and sequential text substitution that cascades or corrupts element symbols. Here we propose TCSP3.0, which enforces coefficient-preserving mappings, substitutes species simultaneously on parsed structures and validates every candidate's composition and cell. We further add periodic-number retrieval, soft oxidation-state evidence, radius-compatibility scoring and structural deduplication. It is further enhanced by adding ORB-v3 machine learning potential based relaxation and energy ranking of its exported candidates, widely used by the deep CSP baselines. TCSP3.0 achieves Top-1 and Top-20 match rates of 72.2\% and 90.6\% on T180, respectively, compared with 28.3\% and 45.6\% for TCSP2.0. On the new 800-target benchmark (T800), its Top-20 match rate is 74.0\%. It also outperforms PNcsp+ and two leading deep learning based generative CSP models, CrystalFlow and MaskGXT, on T180 (Top-1 72.2\% versus at most 63.9\%); even without any relaxation, the native TCSP3.0 achieves 67.8\% at Top-1 and 90.6\% at Top-20 in a median of 28~s per target on two CPU cores, about ten times faster than the complete CrystalFlow job (290~s, including its ORB-v3 stage) on 60 additional targets outside T180. Paired ten-arm ablations of the native chemical-score stage on T180 and T800 show that most of the improvement comes from the correctness fixes while structural deduplication significantly improves Top-5 match rate, embedding-based retrieval is required for coverage, and the remaining scoring features have small or null effects on the match rate.
\end{abstract}

\section{Introduction}
Discovering new inorganic materials underpins progress in several critical industries, from lithium-ion and solid-state batteries for electric vehicles and grid storage to photovoltaics, catalysts for clean fuels, thermoelectrics and semiconductors. A material’s properties are determined by its crystal structure, which fixes its coordination environments, bonding and electronic structure, yet for a compound that has not yet been synthesized this structure is unknown. Crystal structure prediction (CSP), which maps a chemical composition to a periodic arrangement of atoms, is therefore a key step in computational materials discovery. Reliable structure predictions allow promising compositions to be screened for stability and properties before synthesis, narrowing the experimental search and providing starting points for diffraction analysis and first-principles calculations.
Recently, diffusion, flow and autoregressive deep learning models~\cite{DiffCSP2023,EquiCSP2024,CrystalFlow2025,MatterGen2025} have emerged as competitive methods that learn to generate structures, including the lattice parameters and atomic coordinates, directly. Template-based prediction algorithms such as TCSP and CSPML~\cite{TCSP2022,CSPML2022} offer a complementary route when chemically related structures are available, mirroring how chemists reason by analogy (an unknown phase is assumed to adopt the structure type of a chemically similar known compound): they start from a known atomic arrangement of the templates and focus computation on selecting analogues, transferring elemental sites and ranking and relaxing the resulting candidates. These element-substitution based CSP algorithms have demonstrated highly competitive performance~\cite{CSPML2022,Shotgun2024,TCSP2}, which can be evaluated by checking whether a candidate prediction matches a designated reference structure under fixed StructureMatcher settings.

The success of template-based CSP algorithms comes from the fact that many crystal materials reuse related structural prototypes. ICSD~\cite{Allmann2007}, the largest collection of experimentally synthesized crystals, records this recurrence through structure-type assignments: its 2019 account assigns about 159,000 entries, approximately 80\% of the database, to 9,015 structure types shared by at least two compounds~\cite{ICSD2019}. In a template-based CSP algorithm, successful transfer of structures across different chemistries depends on both structural similarity and chemical compatibility. Here the template structure specifies the spatial arrangement of sites, whereas composition and space-group labels provide only part of that information.
The wide sharing of structures by different crystals is also reflected in research on structural prototype catalogues. AFLOW provides crystallographic descriptions for 1,783 structural prototypes in Part 4, with the 2026 Part~5 article adding 344 for a total of 2,127~\cite{AFLOW2024,AFLOWPart5Preprint2026}. CSPD uses structure-similarity analysis to organize 26,023 inorganic prototypes across 8,097 composition types~\cite{CSPD2017}.
Perovskites are a prominent example: about 1,600 compounds have been identified experimentally as perovskites~\cite{Filip2018Perovskite}, and chemically different ones such as the oxides BaTiO$_3$ and SrTiO$_3$ and the halides CsPbI$_3$ and CH$_3$NH$_3$PbI$_3$ all adopt the same ABX$_3$ framework of corner-sharing BX$_6$ octahedra.

One of the key steps in template-based CSP algorithms is finding suitable templates for a given composition. First, chemical priors can help identify such transferable frameworks. An ABX$_3$ perovskite arrangement, for example, supplies characteristic coordination environments that can be screened through charge balance and size compatibility. Statistical ionic substitution learns transfer probabilities from observed compounds~\cite{Hautier2011}. Bartel and colleagues' tolerance factor classifies perovskite occurrence with 92\% accuracy among 576 experimentally characterized oxide and halide ABX$_3$ compounds~\cite{Perovskite2019}. Such priors guide candidate generation, with subsequent physical assessment addressing energy and stability. Following these ideas, TCSP and CSPML established composition-conditioned template prediction~\cite{TCSP2022} and learned substitution similarity~\cite{CSPML2022}. TCSP2.0~\cite{TCSP2} added a large template library, BERTOS oxidation-state inference~\cite{BERTOS2023}, element embeddings and chemistry heuristics. When we re-ran the published TCSP2.0 code on the T180 benchmark, however, it matched only 82 of 180 reference structures within its top 20 candidates. Tracing each of the 2,347 structures that TCSP2.0 predicted across the 180 targets (up to 20 per target) back to its source template showed why: 469 of them (one in five) do not have the requested composition. The code assigns elements without matching stoichiometric coefficients, rewrites CIF text with sequential string replacements that cascade (an inserted element is replaced again) or corrupt element symbols (F replaced inside Fe), and never checks the composition of its output. Another 24 queries crash (23 in an oxidation-state fallback, one in symmetry lookup).

TCSP3.0 fixes these errors and adds further features. Element mappings are restricted to pairs with equal coefficients, substitution acts on all parsed species simultaneously, and each candidate's source composition, output composition and cell size are validated before ranking. On this corrected core, TCSP3.0 adds a periodic-number retrieval route alongside embedding retrieval, multiple near-optimal mappings, probability-weighted oxidation-state evidence, coordination-dependent radius compatibility and structural deduplication of the ranked list. Together these changes double the Top-20 match rate on T180, from 45.6\% to 90.6\%, before any relaxation; ORB-v3~\cite{Rhodes2025Orbv3} relaxation and energy ranking of the exported candidates, the default final stage, then raises the Top-1 match rate to 72.2\%, which beats the two best deep generative CSP models we evaluated on T180, CrystalFlow (Top-1 match rate 63.9\%) and MaskGXT (62.8\%).

Our contributions are: (i) a diagnosis of the original TCSP2.0 implementation that attributes all 469 wrong-composition selections to specific mapping and substitution defects; (ii) TCSP3.0, which corrects these defects and adds the further features described above; (iii) an evaluation on T46, T180 and an independent 800-target cohort (T800), with original TCSP2.0 and PNcsp+ compared on T180 and CrystalFlow and MaskGXT compared on T180 and T46 (T800 is used for TCSP3.0 and its ablations only), using a single StructureMatcher evaluator that keeps every failed target in the denominator~\cite{CSPMetrics2024,CSP1802026,T46Preprint2026}; and (iv) paired ten-arm ablations on T180 and T800 that separate the contribution of each added feature from that of the correctness fixes.

\section{Related work}

There are several important works in the area of template-based crystal structure prediction. Because of their importance, speed and competitive performance, we summarize the major template-based algorithms here.

\subsection{Statistical ionic substitution}
Early data-mining CSP combined correlations among known compounds with quantum-mechanical assessment~\cite{Fischer2006}. Chemical analogy was also used to propose missing ternary oxides for DFT stability evaluation~\cite{Hautier2010}. Within this lineage, ionic substitution learns which oxidation-labelled species can replace one another while preserving a known structural arrangement. Hautier and colleagues fit an ionic-substitution model to observed compounds and used it to propose new compositions and structural analogues~\cite{Hautier2011}. This established data-driven substitution as a candidate-generation strategy: chemical evidence restricts which known arrangements are worth transferring, while subsequent physical assessment addresses their stability. A substitution score is therefore distinct from a calculated energy or a guarantee of a reference match.

\subsection{Contextual and learned chemical similarity}
Another category of template-based CSP algorithms is based on the fact that learned similarity can incorporate more context than a pairwise ionic statistic. Collaborative CSP uses symmetric matrix factorization, optionally with chemical side information~\cite{Hwang2016}. Other models condition compatibility on coordination and topology~\cite{Ryan2018}, while anion-dependent oxidation-state occurrence helps identify plausible chemical analogues~\cite{Davies2018}. Element embeddings supply another substitution-similarity representation~\cite{Onwuli2023}. Together, these methods motivate context-aware template screening.

\subsection{Template-based CSP workflows: CSPML, ShotgunCSP and TCSP}

There are several ways to select templates and conduct element substitution.
CSPML~\cite{CSPML2022} selects composition-ratio-compatible templates using learned pair structural similarity. It preserves coefficients during substitution and resolves ambiguous equal-coefficient assignments through descriptor distances before DFT relaxation. Its reported 96.4\% pair-classification accuracy measures template similarity. This establishes coefficient-preserving transfer and descriptor-based assignment as prior methods.

ShotgunCSP combines a template-based generator (GT) with a template-free Wyckoff generator (GW)~\cite{Shotgun2024,ShotgunCorrection2025}. GT uses equal-coefficient substitution, descriptor assignment, composition clustering and StructureMatcher deduplication, followed by lattice and coordinate perturbations. Target-specific CGCNN screening of DFT single points prioritizes candidates for final DFT refinement. Coordinate-free coarse graining also uses Wyckoff information to prioritize stability~\cite{Goodall2022}. These approaches couple structural generation to a separate physical-assessment budget.

TCSP uses composition-conditioned template substitution~\cite{TCSP2022}. TCSP2.0 adds BERTOS oxidation-state prediction, element embeddings and chemistry-based ranking~\cite{TCSP2,BERTOS2023}. In this study we evaluate the unchanged original TCSP2.0 implementation on T180, preserving its mapping, relaxation and final-selection behavior on the same template library (the query-exclusion rules differ; see the ``Comparison with the original TCSP2.0 implementation'' section), and separately a repaired internal baseline used in the ablation (the ``Ablation Study'' section).

\subsection{Periodic-number retrieval and PNcsp+}
PNcsp uses periodic-number crystal-structure similarity~\cite{PNcsp2024} for template selection. PNcsp+ adds learned-potential evaluation~\cite{PNcsp2026}. PNcsp+ retrieves OQMD structures through periodic-number chemical neighbourhoods and ranks substituted candidates by energy~\cite{PNcsp2026,PNcspSoftware2026}. Its benchmark searches order-four neighbours, extending to six when no candidate is found, and averages unrelaxed MACE MPA-0 and M3GNet energies per atom. TCSP3.0 adapts PN neighbourhoods as a retrieval route alongside embedding retrieval. Our direct PNcsp+ comparison (the ``Comparison with PNcsp+'' section) keeps its retrieval and ranking rules but uses the LeMat-derived OQMD inventory and omits the negative-formation-energy filter.

\subsection{Comparison axes and complementary coordinate generation}
The above-mentioned template-based CSP methods address different stages of a template workflow (Table~\ref{tab:lineage} in the Appendix). Geometric preparation of the substituted template, such as reference-based or data-mined lattice scaling, is addressed separately~\cite{Chu2018}.

On the other hand, deep generative models such as DiffCSP, EquiCSP, FlowMM and CrystalFlow learn distributions of crystal structures, including their lattices and coordinates~\cite{DiffCSP2023,EquiCSP2024,FlowMM2024,CrystalFlow2025}. MatterGen extends this approach to generative materials design~\cite{MatterGen2025}. Our comparison evaluates archived generator predictions on the same T46/T180 reference cohorts, with their training, conditioning and relaxation settings documented.

\section{Methods}
\subsection{Overview of the TCSP3.0 algorithm}
\begin{figure}[!htbp]
\centering\includegraphics[width=\linewidth]{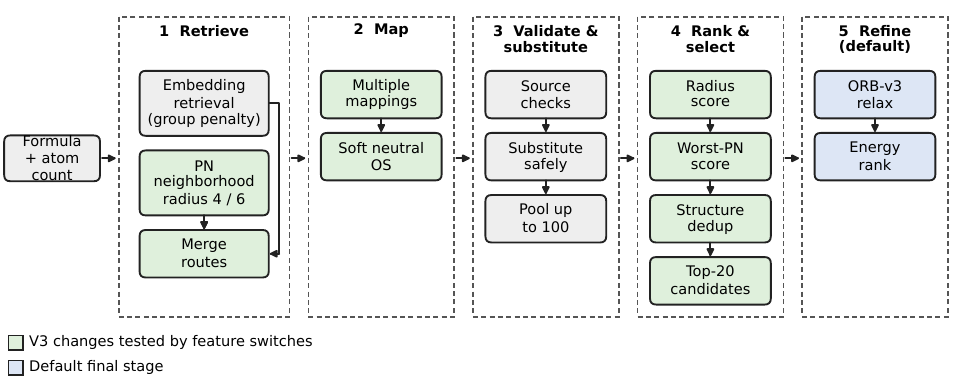}
\caption{TCSP3.0 workflow. Retrieval and mapping use composition alone; geometry comes from a source template. Up to 100 valid candidates are reranked by chemical and geometric scores and deduplicated. Green boxes mark V3 changes tested by the feature switches of the ``Ablation Study'' section, and the blue boxes are the default final stage; the correctness safeguards in the grey source-check box are shared by all configurations. Reference structures are used only in evaluation.}
\label{fig:workflow}
\end{figure}

TCSP3.0 predicts crystal structures by transferring known templates to a target composition through coefficient-preserving element substitution. Figure~\ref{fig:workflow} summarizes the workflow, which takes only the integer composition and output-cell atom count as inputs. Embedding and periodic-number retrieval identify templates; mapping uses soft oxidation-state evidence, followed by source validation and simultaneous species substitution. The valid candidate pool is ranked by chemical and radius-compatibility scores, then structurally deduplicated and exported. By default, ORB-v3 relaxes the exported candidates and ranks them by energy per atom (the ``ORB-v3 relaxation and ranking (default final stage)'' section); feature ablations instead use the unrelaxed native chemical-score order. Reference coordinates, space group and energy are used only for evaluation. The method reuses the TCSP2.0 template library, Matscholar element embeddings~\cite{Tshitoyan2019} and pretrained BERTOS checkpoint~\cite{BERTOS2023}, without training a model or generating coordinates de novo. The template library is the TCSP2.0 index of about 731K entries from which the MP-20 test structures are removed, leaving 719,450 entries (659,589 distinct formulas).
 Template generation uses the frozen source commit \texttt{9fd44314}~\cite{TCSPV3Software2026}; the ORB-v3 default final stage is provided in release 3.1.0 (commit \texttt{d8383d3}).

\subsection{Diagnosis of TCSP2.0}\label{sec:tcsp2-failure}


We analyzed the 2,347 predictions of TCSP2.0 on the T180 dataset with top-20 results, of which 469 had incorrect compositions already present before CHGNet relaxation. Element assignments do not enforce stoichiometric coefficients. Sequential CIF-text replacement can re-substitute newly inserted elements or rewrite one element symbol inside another, such as F inside Fe. Without output-composition validation, these candidates remain eligible for selection (Supplementary Section~\siref{sec:original-diagnosis}).
To address these issues, TCSP3.0 applies coefficient-preserving mappings, simultaneous species substitution and source/output-composition and cell-count validation in its frozen template-generation stage.

\subsection{Complementary retrieval and coefficient-preserving mappings}\label{sec:complementary-retrieval}
In TCSP3.0, the query composition is first reduced to positive integer stoichiometric coefficients. Templates must share the sorted coefficient signature, and the query's own reduced composition is excluded. The embedding route ranks template formulas by two distances, an electronegativity-ordered distance and an assignment-based distance, and keeps up to 200 formulas from each (at most 400 in total). Element distance is the Euclidean distance between Matscholar embeddings plus a unit penalty for different periodic groups; elements without embeddings fall back to periodic-number (PN) distance.

The second route requires every mapped element pair to lie within a Villars PN radius of four, extended to six only when no formula qualifies at four, and keeps up to 200 formulas. The formulas retrieved by the embedding and PN routes are interleaved into one ordered list, counting duplicate formula/mapping combinations once.

For each query/template pair, the Hungarian algorithm~\cite{Kuhn1955} finds the minimum-cost assignment under the additive embedding and oxidation-state cost, with edges between elements of different coefficients forbidden. TCSP3.0 additionally enumerates permutations within equal-coefficient groups (at most 40,320), orders them by the full score $C$ of Eq.~\eqref{eq:score}, which includes the worst-pair PN penalty, and keeps up to three mappings within 0.25 of the best. The one-mapping ablation keeps only the Hungarian assignment, which can differ from the best full-score mapping.

For mapping $\pi$, the implemented score can be written as
\begin{align}\label{eq:score}
C(\pi)&=\sum_i\frac{n_i}{N}\left[d(e_i,t_{\pi(i)})+
0.5\,m\sum_{a,b}w_a w_b\min\left(\frac{|o_{i,a}-o_{\pi(i),b}|}{2},2\right)\right]\nonumber\\
&\quad+0.1\min\left(\frac{\max_i|\mathrm{PN}(e_i)-\mathrm{PN}(t_{\pi(i)})|}{6},2\right),
\end{align}
where $n_i/N$ is the atomic fraction of element $i$ in the reduced formula, $m$ is the smaller neutral oxidation-evidence mass of query and template, $w_a,w_b$ are neutral-assignment weights, and $o_{i,a}$, $o_{\pi(i),b}$ are the oxidation states of the neutral assignments of query element $i$ and template element $\pi(i)$. Oxidation differences are normalized by two charge units and PN differences by six steps, each capped at 2. The weights 0.5 and 0.1 were set by hand, not tuned on any benchmark, and held fixed throughout.

\subsection{Soft oxidation evidence}
TCSP3.0 uses the pretrained BERTOS predictor~\cite{BERTOS2023} unchanged but treats its outputs as soft evidence. Species probabilities are averaged over tokens, the three most probable labels per species are combined, and up to four charge-neutral assignments are kept, weighted by the product of their species probabilities. These weights enter the mismatch term of Eq.~\eqref{eq:score}, scaled by the total neutral probability mass. Compositions with no neutral assignment, including all-metal compositions, receive no oxidation penalty. The single-OS ablation keeps only the first (most probable) neutral assignment at the same total mass.

\subsection{Source validation and geometric compatibility}\label{sec:source-validation}
Before element substitution, each source's actual composition is checked against the index and against the excluded query composition, and disordered sources are rejected. All species are substituted simultaneously on the parsed structure, so an inserted element is never replaced again, and the resulting reduced composition must equal the query. In full-formula mode the source cell must also contain the requested number of atoms, and sources with interatomic contacts below 0.5~\AA{} are rejected. These correctness safeguards, with implementation details in Supplementary Section~\siref{sec:correctness}, are shared by all ten ablation configurations.

In TCSP3.0, nearest-shell coordination counts neighbors within 1.25 times the nearest distance, using a 5~\AA{} search radius. Compatible Shannon radii are used when oxidation and coordination are available for both elements; otherwise both use atomic radii. Ionic and atomic radii are not mixed in one comparison. The geometric penalty is
\begin{equation}
G=0.25\,\mathrm{mean}\left|\log(r_{\mathrm{query}}/r_{\mathrm{template}})\right|.
\end{equation}
This penalty reranks the capped valid pool.

\subsection{Selection and budgets}
The pool holds up to 100 valid candidates, which may come from different templates or mappings. A formula/mapping row is one template formula paired with one retained element mapping; rows follow the interleaved order of the embedding and PN routes. Each row is expanded completely, with its source structures taken in lexicographic ID order, before the next row, until 100 valid candidates have been admitted.
The pool is sorted by $C+G$, and up to 20 structures are exported after greedy pairwise StructureMatcher deduplication~\cite{Pymatgen2013}, which compares candidates across space-group labels. Targets with fewer than 20 distinct candidates export fewer. The no-dedup ablation exports the top 20 pool entries without deduplication, so its 20 entries need not be distinct.

\subsection{Datasets}\label{sec:datasets}
We constructed revised T46 and T180 benchmarks to improve reference quality and control overlap with model-training data. The earlier T46 set, although selected to exclude MP-20 training and validation entries, still contained targets requiring correction; the original T180 set also overlapped those splits. Phonon-unstable references in the MP-20 test split further motivated dynamical-stability checks alongside thermodynamic screening by energy above the convex hull (below 0.1~eV/atom for all targets). Purely theoretical references were required to be dynamically stable, whereas experimentally known compounds were retained regardless of their DFT phonon results; the five T180 references with imaginary modes are listed in Supplementary Section~\siref{sec:benchmark-inventories}.

We use the revised T46 and T180 benchmarks (release v3.1; MP-ids in the supplementary file). They share 33 targets, giving 193 distinct targets; shared targets are reported in both cohorts. Complete target inventories and released metadata are provided in Supplementary Tables~\siref{tab:t46_inventory} and~\siref{tab:t180_inventory}. Ablations are run on T180 and on the T800 cohort described below.
Run-specific input-cell details are given in Supplementary Section~\siref{sec:benchmark-inventories}.
None of the 193 reference structures matches any of the 36,183 MP-20 training/validation structures under StructureMatcher at $(0.2,0.3,5^\circ)$ or $(0.3,0.5,10^\circ)$ tolerances. Reusing a chemically different known prototype is intrinsic to template prediction and is permitted.

The T800 cohort contains 800 targets with no ID, composition or structural overlap with T46/T180. It is drawn almost entirely from the MPTS-52 test split (796 test, two training and two validation records). Four queries (\texttt{mp-1205423}, \texttt{mp-24971}, \texttt{mp-1190827}, \texttt{mp-1202027}) use their released 24/9/24/24-atom formulas although their spglib cells have 48/18/48/48 atoms. T800 targets contain 4--52 atoms, compared with 2--28 in T180. The cohorts also differ in chemistry: oxygen-containing compounds are 22.2\% of T180 but 51.0\% of T800, and intermetallic compounds 25.6\% and 10.4\% (Supplementary Table~\siref{tab:composition}).
T800 evaluation uses the same code, ten configurations and 719,450-entry template index, with each query's own composition and structurally equivalent aliases excluded. Other T800 compositions, including other benchmark targets, remain eligible as templates; no global cohort or prototype exclusion is applied. Re-running the T180 configuration reproduced every pool, rank, score and metric before T800 generation. The T800 candidates of the full configuration were then relaxed and ranked with ORB-v3 in the same way (781 targets have candidates; the other 19 count as misses).

\subsection{Evaluation and uncertainty}
The Top-$k$ match rate, widely used in CSP evaluation, is the fraction of targets for which at least one of the first $k$ ranked candidates matches the reference under pymatgen StructureMatcher (\texttt{ltol=0.2}, \texttt{stol=0.3}, angle tolerance $5^\circ$, primitive reduction and scaling enabled, supercell and subset matching disabled). Every target stays in the denominator; failed generations and invalid candidates count as misses.

We also report counts with 95\% percentile bootstrap intervals from 10,000 target resamples (seed 20261010). Feature contrasts use paired resampling and exact two-sided McNemar tests on gained and lost targets, with Holm correction over the nine ablation contrasts within each cohort and metric. TCSP3.0 is deterministic.
Space-group (SG) labels of predictions and references are determined independently at symprec 0.1~\AA{} and angle tolerance $5^\circ$ (T800 also at 0.01~\AA{}). Consensus requires the same candidate to match both structure and SG.

\subsection{ORB-v3 relaxation and ranking (default final stage)}\label{sec:orb-default}
In TCSP3.0 predictions, the exported candidates of each target (at most 20, fewer when fewer distinct structures exist) are relaxed with ORB-v3 (\texttt{orb-v3-conservative-inf-omat}) using FIRE with ASE \texttt{ExpCellFilter}, a 0.01~eV/\AA{} force threshold and at most 500 steps, and are ranked by ORB-v3 energy per atom. This is also the post-processing applied to the deep generative baselines (the ``Comparison with deep generative CSP models'' section), so TCSP3.0 and the generators are compared with the same final stage. Unless a result is labelled native, TCSP3.0 denotes this default configuration. All candidates relaxed without failure, nonconverged structures were kept (68.8\% of the 2,581 T180 candidates reached the force threshold), and every target stays in the denominator. The native configuration, which keeps the chemical-score order and applies no relaxation, is the reference for the feature ablations and for the refinement analyses in Supplementary Sections~\siref{sec:mattersim-refinement} and~\siref{sec:orb-refinement}, where the refined candidates are compared with it.

\subsection{MatterSim refinement}
As an alternative machine learning potential we also relaxed all 2,581 exported T180 candidates with MatterSim (\texttt{MatterSim-v1.0.0-1M})~\cite{MatterSim2024}, using FIRE with ASE \texttt{ExpCellFilter} on atomic positions and all cell degrees of freedom, a 0.01~eV/\AA{} force threshold and at most 500 steps.
We evaluate two orderings: relaxed geometries in the original chemical-rank order, and the same relaxed candidates sorted by MatterSim energy per atom. Both keep nonconverged structures and all 180 targets.

\section{Results}
\subsection{Overall performance}
\begin{figure}[!htbp]\centering
\includegraphics[width=\linewidth]{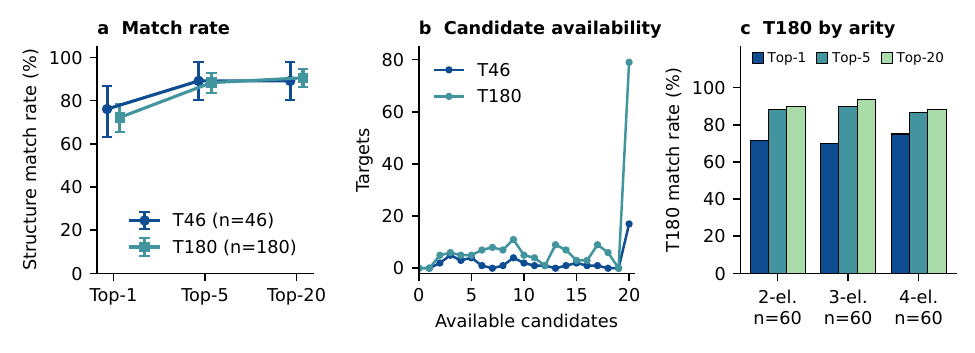}
\caption{TCSP3.0 performance. (a) Top-$k$ match rate with 95\% bootstrap intervals. (b) Number of candidates per target. (c) T180 match rate by number of elements (60 targets each).}
\label{fig:current}\end{figure}

Table~\ref{tab:current} reports match rates of TCSP3.0 in its default configuration (ORB-v3 relaxation and energy ranking) on T46, T180 and T800 (Figure~\ref{fig:current}a shows T46 and T180); the native chemical order, without relaxation, gives 34/39/41 on T46, 122/154/163 on T180 and 512/575/591 on T800 (Supplementary Section~\siref{sec:orb-refinement}). Every exported CIF (2,581 for T180, 561 for T46) has the requested composition and cell size. Every T46 and T180 target receives at least one candidate, but only 79/180 T180 and 17/46 T46 targets yield the full 20 distinct candidates (Figure~\ref{fig:current}b). The match rate does not fall with the number of elements: binary, ternary and quaternary T180 targets (60 each) reach 43, 42 and 45 matches at Top-1 and 54, 56 and 53 at Top-20 (Figure~\ref{fig:current}c). The overall rate is therefore not driven by the simpler binary compounds.

\begin{table}[!htbp]\centering
\caption{Top-$k$ structure match rate of TCSP3.0 with ORB-v3 relaxation and energy ranking. Brackets are 95\% bootstrap intervals. The disjoint subset removes the 33 targets shared with T46.}
\begin{tabular}{lrrr}\toprule
Cohort & Top-1 & Top-5 & Top-20\\\midrule
T46 ($n=46$) & 35 (76.1\%) & 41 (89.1\%) & 41 (89.1\%)\\
95\% interval & [63.0,87.0] & [80.4,97.8] & [80.4,97.8]\\
T180 ($n=180$) & 130 (72.2\%) & 159 (88.3\%) & 163 (90.6\%)\\
95\% interval & [65.6,78.3] & [83.3,92.8] & [86.1,94.4]\\
T180 disjoint ($n=147$) & 103 (70.1\%) & 129 (87.8\%) & 133 (90.5\%)\\
95\% interval & [62.6,77.6] & [82.3,92.5] & [85.7,95.2]\\
T800 ($n=800$) & 518 (64.8\%) & 579 (72.4\%) & 592 (74.0\%)\\
95\% interval & [61.4,68.0] & [69.2,75.5] & [71.0,77.0]\\\bottomrule
\end{tabular}\label{tab:current}
\end{table}

On T180, moving from Top-1 to Top-5 adds 29 matched targets, and ranks 6--20 add four more (six and none on T46). Seventeen T180 and five T46 targets have no match among their candidates.

The following subsections compare TCSP3.0 with TCSP2.0 (the ``Comparison with the original TCSP2.0 implementation'' section), PNcsp+ (the ``Comparison with PNcsp+'' section) and two deep generative models (the ``Comparison with deep generative CSP models'' section), ablate its components (the ``Ablation Study'' section), examine refinement (the ``Learned-potential refinement'' section), analyze the remaining failures (the ``Failure analysis on T180'' section) and report recovery by chemical family (the ``Performance by chemical family'' section).

\FloatBarrier
\subsection{Comparison with the original TCSP2.0 implementation}\label{sec:cmp2}
We first compare TCSP3.0 with the original TCSP2.0 code~\cite{TCSP2}, run on the same 180 T180 queries with its BERTOS checkpoint~\cite{BERTOS2023}, element substitution, CHGNet relaxation~\cite{CHGNet2023} and chemical-score selection. The original code was run on T180 only; the T180 disjoint subset ($n=147$; Table~\ref{tab:current}) removes the 33 targets shared with T46 so that no target is counted in both cohorts.
Its template library was restricted to the 719,450 IDs eligible for TCSP3.0. For TCSP2.0, only each query's own ID and structurally equivalent aliases (53 entries over 44 queries) were removed, so same-composition polymorphs remained eligible; for TCSP3.0 all templates with the query's reduced composition are excluded, which is the stricter setting.
We raised the formula shortlist from the default of five to 20 so that up to 20 candidates are returned. CHGNet 0.3.0 relaxes each candidate (FIRE, cell relaxation, 0.1~eV/\AA{}, at most 500 steps).

\begin{figure}[!htbp]\centering
\includegraphics[width=\linewidth]{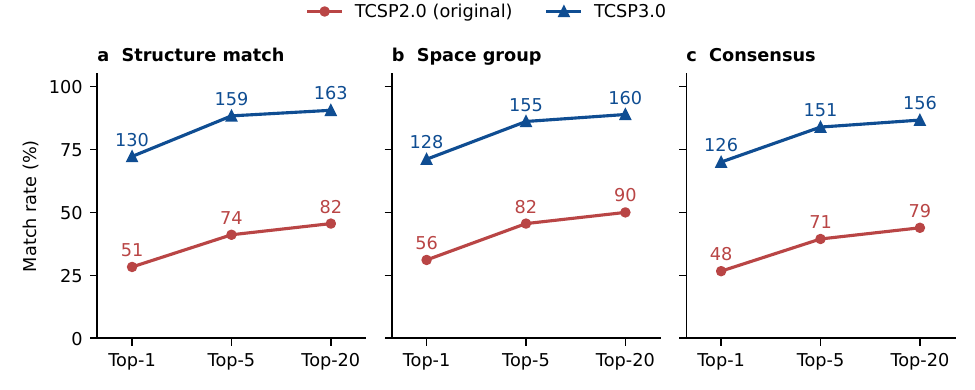}
\caption{Match rate of the original TCSP2.0 code and TCSP3.0 (ORB-v3 relaxed and ranked) on the 180 T180 targets. Numbers give matched targets; failed runs and wrong-composition candidates count as misses.}\label{fig:tcsp2_three_methods}
\end{figure}

Table~\ref{tab:tcsp2context} and Figure~\ref{fig:tcsp2_three_methods} compare the two implementations.\footnote{A comparator that repairs only the two crashing code paths of TCSP2.0 (crashfix6/7), leaving all mapping and substitution defects intact, matches 52/75/83 targets at Top-1/5/20, one more than the original at each cutoff (Supplementary Section~\siref{sec:crashfix67}).}

\begin{table}[!htbp]\centering\small
\setlength{\tabcolsep}{4pt}
\caption{Counts and match rates on all 180 T180 targets for the original TCSP2.0 code and TCSP3.0. Cells give matched-target counts with percentages in parentheses (denominator 180). TCSP3.0 is shown in its default configuration (ORB-v3 relaxation and energy ranking) and with its candidates in the native chemical order, which isolates the stage before relaxation and is the closer counterpart of the chemical-score selection of TCSP2.0. Failed runs and wrong-composition candidates count as misses. ``(CPU)'' denotes the original code as run on CPU hardware; the crash-repaired comparator was run on different hardware (Supplementary Section~\siref{sec:crashfix67}). SG is computed from each output CIF at 0.1~\AA{}; consensus requires one candidate to match both structure and SG.}
\begin{tabularx}{\linewidth}{>{\raggedright\arraybackslash}Xlrrr}\toprule
Implementation & Criterion & Top-1 & Top-5 & Top-20\\\midrule
Original TCSP2.0 (CPU) & Structure & 51 (28.3\%) & 74 (41.1\%) & 82 (45.6\%)\\
 & SG & 56 (31.1\%) & 82 (45.6\%) & 90 (50.0\%)\\
 & Consensus & 48 (26.7\%) & 71 (39.4\%) & 79 (43.9\%)\\
TCSP3.0 (ORB-v3 relax and rank) & Structure & 130 (72.2\%) & 159 (88.3\%) & 163 (90.6\%)\\
 & SG & 128 (71.1\%) & 155 (86.1\%) & 160 (88.9\%)\\
 & Consensus & 126 (70.0\%) & 151 (83.9\%) & 156 (86.7\%)\\
TCSP3.0 native order (no relaxation) & Structure & 122 (67.8\%) & 154 (85.6\%) & 163 (90.6\%)\\
 & SG & 125 (69.4\%) & 152 (84.4\%) & 160 (88.9\%)\\
 & Consensus & 117 (65.0\%) & 147 (81.7\%) & 156 (86.7\%)\\\bottomrule
\end{tabularx}\label{tab:tcsp2context}
\end{table}

TCSP3.0 doubles Top-20 structure match rate, from 82/180 (45.6\%) to 163/180 (90.6\%), gaining 82 targets and losing one ($+45.0$ points, 95\% interval $[37.2,52.2]$; Holm-adjusted McNemar $p=1.6\times10^{-22}$), and raises Top-1 from 51 to 130 targets. The gain holds on the T180 disjoint subset ($n=147$; 64 versus 133).
The original code also fails outright on 24 of the 180 queries: 23 crash on an undefined \texttt{elements} variable in the oxidation-state fallback, and one stops during symmetry lookup.

Beyond these failures, 469 of the 2,347 selected CIFs (20\%) by TCSP2.0 have the wrong composition, although their site counts are correct. Tracing each to its source template identifies the three defects of the ``Diagnosis of TCSP2.0'' section (Supplementary Section~\siref{sec:original-diagnosis}), and no output-composition check catches them. TCSP3.0 eliminates all three by construction (the ``Source validation and geometric compatibility'' section); the ablations in the ``Ablation Study'' section quantify how much of the gain they explain.

\FloatBarrier
\subsection{Comparison with PNcsp+}\label{sec:pncsp}

The closest template-based competitor is PNcsp+, which differs from TCSP3.0 in its template inventory, retrieval and ranking. Both are evaluated on T180 with the same procedure. To ensure fairness, we adapt PNcsp+ to draw templates from the 1,076,924 OQMD-derived structures in LeMat. This export lacks OQMD formation energies, so we omit the published negative-formation-energy filter; otherwise we keep the official order-four PN retrieval with order-six fallback, upstream substitution and ranking by the mean unrelaxed MACE MPA-0 and M3GNet energy, with no candidate cap.

PNcsp+ matches 71/96/108 targets at Top-1/5/20 from its full energy ranking and 71/106/108 from its list keeping the best candidate per source space group (Table~\ref{tab:pncsp_current}). 29 targets receive no candidate, of which 19 have no source within PN order six and ten fail an upstream check that compares reduced candidate counts with the literal input formula.

\begin{table}[!htbp]\centering\small
\caption{T180 comparison of TCSP3.0 and PNcsp+ under the common reference evaluator. Cells show matched targets (percent), with all 180 targets retained. PNcsp+ uses the LeMat-derived OQMD adaptation and official supported no-formation-energy-filter mode; SG-best retains one candidate per derived source-SG label. Template eligibility, cell sizes and candidate/physical budgets differ.}\label{tab:pncsp_current}
\begin{tabular}{llrrr}\toprule
Metric & Rank & TCSP3.0 (ORB-v3) & PNcsp+ full & PNcsp+ SG-best\\\midrule
Structure & Top-1 & 130 (72.2\%) & 71 (39.4\%) & 71 (39.4\%)\\
Structure & Top-5 & 159 (88.3\%) & 96 (53.3\%) & 106 (58.9\%)\\
Structure & Top-20 & 163 (90.6\%) & 108 (60.0\%) & 108 (60.0\%)\\
\midrule
Space group & Top-1 & 128 (71.1\%) & 68 (37.8\%) & 68 (37.8\%)\\
Space group & Top-5 & 155 (86.1\%) & 93 (51.7\%) & 110 (61.1\%)\\
Space group & Top-20 & 160 (88.9\%) & 111 (61.7\%) & 118 (65.6\%)\\
\midrule
Consensus & Top-1 & 126 (70.0\%) & 68 (37.8\%) & 68 (37.8\%)\\
Consensus & Top-5 & 151 (83.9\%) & 90 (50.0\%) & 101 (56.1\%)\\
Consensus & Top-20 & 156 (86.7\%) & 104 (57.8\%) & 103 (57.2\%)\\
\bottomrule\end{tabular}
\end{table}

\begin{figure}[!htbp]\centering
\includegraphics[width=0.97\linewidth]{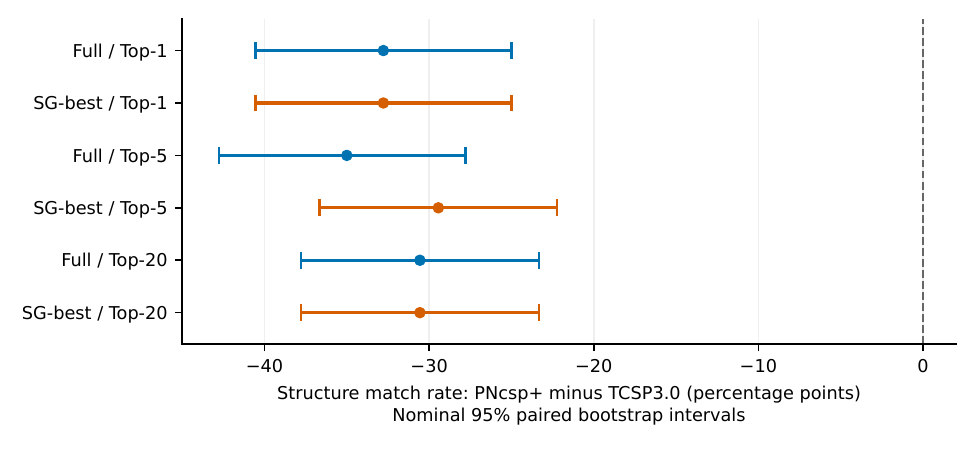}
\caption{Paired structure match-rate differences of PNcsp+ relative to TCSP3.0 on all 180 T180 targets, with nominal 95\% paired bootstrap intervals; negative values favor TCSP3.0. Full is the full energy ranking and SG-best the SG-best list of Table~\ref{tab:pncsp_current}.}\label{fig:pncsp_paired}
\end{figure}

At Top-5, TCSP3.0 matches 67 targets that the full PNcsp+ ranking misses and misses four that it matches (PNcsp+ minus TCSP3.0: $-35.0$ points, 95\% interval $[-42.8,-27.8]$); against the SG-best list the counts are 58 and five ($-29.4$ points, $[-36.7,-22.2]$; Figure~\ref{fig:pncsp_paired}).
On the T180 disjoint subset ($n=147$), full and SG-best match 47/68/77 and 47/76/78 references, respectively, compared with TCSP3.0's 103/129/133.

\FloatBarrier
\subsection{Comparison with deep generative CSP models}\label{sec:deepcmp}
Finally, we compare TCSP3.0 with the two deep generative models that reach the highest T180 Top-1 match rates among those we evaluated, CrystalFlow~\cite{CrystalFlow2025} and MaskGXT~\cite{Seong2026MaskGXT}, using the same matcher.

The generators were trained on MP-20 and their candidates are relaxed and ranked with ORB-v3, except for MaskGXT, which uses ORB-v3 single-point ranking without relaxation. TCSP3.0 receives the same ORB-v3 relaxation and energy ranking, so all three methods share the final stage; the native chemical-score order is reported in Supplementary Section~\siref{sec:orb-refinement}. Counts use all 180 T180 or 46 T46 targets and missing predictions are misses; MaskGXT returns candidates for 176 of 180 T180 targets and 45 of 46 T46 targets. These results come from the original runs of each method, not from matched-compute reruns.

TCSP3.0 matches 130/180 targets at Top-1 (72.2\%), compared with 115/180 (63.9\%) for CrystalFlow and 113/180 (62.8\%) for MaskGXT. The margin grows at Top-5 (159 versus 121 and 116) and Top-20 (163 versus 130 and 119). On T46, which shares 33 targets with T180, TCSP3.0 reaches 35/41/41, compared with 26/27/31 for CrystalFlow and 30/30/31 for MaskGXT. Space-group agreement of the Top-1 candidate at 0.1~\AA{} (SG1) is 128/180 for TCSP3.0, 112 for CrystalFlow and 72 for MaskGXT, and the corresponding same-candidate consensus of structure and space group is 126, 106 and 68. MaskGXT candidates are unrelaxed, so its SG1 (72 on T180 and 21 on T46) is evaluated on the unrelaxed generated structures and is much lower than its Top-1 (113 and 30). Even without relaxation, the native TCSP3.0 matches 122 T180 targets at Top-1, more than the relaxed CrystalFlow and MaskGXT.

The template stage is also much cheaper than generation. On 60 additional targets run on identical hardware, it takes a median of 28~s per target on two CPU cores (no relaxation), against 290~s for CrystalFlow, whose time includes its ORB-v3 stage. The separate ORB-v3 stage of TCSP3.0 adds about 219~s per target on a GPU, so the full default pipeline costs about as much as CrystalFlow (Supplementary Section~\siref{sec:runtime}).

\FloatBarrier
\subsection{Ablation Study}\label{sec:ablation}
The comparisons above do not show which components matter. We therefore tested the ten configurations of Table~\ref{tab:matrix} on T180 and T800 in the native order, without relaxation, so that retrieval and scoring effects are not mixed with those of the energy model; inputs and template exclusion are identical. Detailed results are in Supplementary Section~\siref{sec:ablation-details}.

\begin{table}[!htbp]\centering\small
\caption{Ablation configurations. All share the same template index, inputs, query-composition exclusion, 100-candidate cap, matcher and correctness safeguards; none uses relaxation.}
\begin{tabularx}{\linewidth}{lX}\toprule
Arm & Change relative to full V3\\\midrule
Full V3 & Native configuration (all V3 features, no relaxation); explicit configuration checked for default parity\\
Embedding only & Remove PN retrieval route; retain V3 scoring and mapping\\
PN only & Remove embedding route; retain V3 scoring and mapping\\
One mapping & Single Hungarian base-cost assignment; combined-score choice can differ\\
Single OS & Retain first (most probable) neutral assignment; evidence mass held fixed\\
No OS & Remove OS assignment-scoring term; geometry evidence unchanged\\
No radius & Remove radius penalty; geometric validity checks retained\\
No worst PN & Remove worst-pair PN score; neighborhood constraints retained\\
No dedup & Export first 20 pool entries without structure removal\\
Internal baseline & Existing repaired mode; not an exact TCSP2.0 reproduction\\\bottomrule
\end{tabularx}\label{tab:matrix}\end{table}

\begin{table}[!htbp]\centering\small
\caption{Controlled feature outcomes on T180 (180 targets); Full V3 is the native TCSP3.0 configuration. Entries are StructureMatcher reference-match counts; every arm retains the full denominator. Pool oracle means at least one valid candidate in the capped pool matches the reference. Top-5 gains/losses compare the ablation with full V3 on paired targets.}
\begin{tabular}{lrrrrr}\toprule
Arm & Top-1 & Top-5 & Top-20 & Pool oracle & Top-5 $+/-$\\\midrule
Full V3 & 122 & 154 & 163 & 165 & --\\
Embedding only & 122 & 154 & 163 & 165 & 0/0\\
PN only & 118 & 148 & 154 & 154 & 1/7\\
One mapping & 122 & 154 & 163 & 165 & 0/0\\
Single OS & 122 & 154 & 163 & 165 & 0/0\\
No OS & 123 & 153 & 162 & 165 & 0/1\\
No radius & 120 & 153 & 163 & 165 & 1/2\\
No worst PN & 121 & 152 & 162 & 165 & 0/2\\
No dedup & 122 & 145 & 159 & 165 & 0/9\\
Internal baseline & 120 & 146 & 161 & 165 & 0/8\\
\bottomrule\end{tabular}\label{tab:controlled_outcomes}\end{table}

Structural deduplication gives the clearest benefit. Removing it leaves the matching pools and Top-1 unchanged but lowers Top-5 structure match rate from 154 to 145 on T180 (Table~\ref{tab:controlled_outcomes}; zero gains, nine losses; Holm-adjusted $p=0.035$) and from 575 to 564 on T800 (one gain, 12 losses; $p=0.027$).

Embedding retrieval is essential for coverage. PN-only retrieval lowers T180 pool-oracle match rate from 165 to 154 (adjusted $p=0.009$) and T800 Top-20 match rate from 591 to 486 (105 losses, no gains; $-13.1$ points; $p=4.4\times10^{-31}$). Embedding-only retrieval, in contrast, reproduces every T180 outcome and every T800 Top-20 outcome: adding the PN route changes the candidate lists but does not improve the match rate.

\begin{figure}[!htbp]\centering\includegraphics[width=\linewidth]{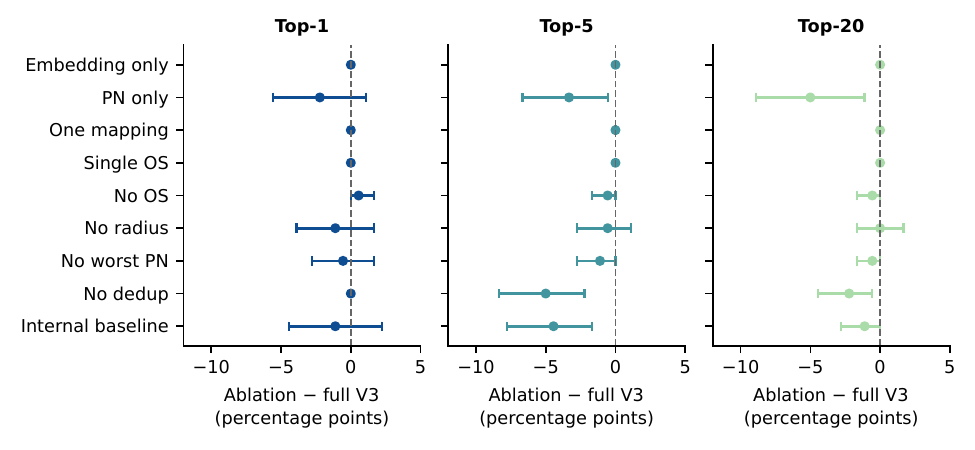}
\caption{Paired ablation-minus-full differences on T180 with 95\% bootstrap intervals; negative values favor the full configuration.}\label{fig:controlled-paired}\end{figure}

The mapping and scoring features have small or null effects. One-mapping and single-OS reproduce every T180 match-rate outcome, and together with no-OS and no-radius they leave T800 Top-20 outcomes unchanged, although each changes candidate scores and orderings. No other scoring contrast reaches adjusted $p<0.05$ (Figure~\ref{fig:controlled-paired}). The repaired internal baseline, which shares the correctness fixes, reaches 161/180 at Top-20, close to the full configuration's 163. The large gain over original TCSP2.0 (the ``Comparison with the original TCSP2.0 implementation'' section) therefore comes mainly from the correctness fixes.

Symmetry metrics show a trade-off. On T800, removing deduplication raises Top-20 structure-and-SG consensus from 542 to 563 (27 gains, six losses; adjusted $p=0.002$), and the internal baseline reaches 570 ($p=4.6\times10^{-7}$), even though both lose structure match rate at Top-5 (Supplementary Tables~\siref{tab:t800_performance} and~\siref{tab:t800_paired}).

\FloatBarrier
\subsection{Learned-potential refinement}\label{sec:refine}
The ablations use the native order, in which the full configuration matches 163/180 targets at Top-20 on T180 and 591/800 (73.9\%) on T800; T800 targets are larger, contain more elements and are richer in oxides (the ``Datasets'' section). We now ask what the final stage adds. MatterSim relaxation with energy re-ranking raises T180 Top-1 match rate from 122 to 126 but lowers Top-20 from 163 to 161 (Supplementary Section~\siref{sec:mattersim-refinement}).

ORB-v3 relaxation with ORB-v3 energy ranking, the default final stage, raises T180 Top-1 structure match rate from 122 to 130 (27 targets gained, 19 lost) and Top-5 from 154 to 159, leaves Top-20 at 163, and raises consensus Top-1 from 117 to 126. Relaxation in the native order changes Top-1 by a single target, so on T180 the gain comes from the energy ranking; on T800, relaxation alone already raises Top-1 from 512 to 518 (Supplementary Table~\siref{tab:orb_current}). The direction is the same on the other cohorts: the native-to-default match counts are 34/39/41 to 35/41/41 on T46 and 512/575/591 to 518/579/592 on T800, where space-group Top-5 rises from 546 to 566 (Supplementary Section~\siref{sec:orb-refinement}).

\FloatBarrier
\subsection{Failure analysis on T180}\label{sec:failure}
We next examine where TCSP3.0 still fails on T180.
TCSP3.0 misses 50 of the 180 T180 targets at Top-1. For 33 of them the reference is among the exported candidates but not ranked first (ranking errors; 29 at rank 2--5, four at rank 6--20), and for 17 it is not among the exported candidates (retrieval failures; Supplementary Section~\siref{sec:failure-analysis}). In the separate MatterSim diagnostic cascade (Supplementary Table~\siref{tab:failure_t180}), the unrelaxed Top-1 candidate relaxes at least 5~meV/atom below the reference energy (F8) for 19 of these 50 misses (18 ranking errors and one retrieval failure, \texttt{mp-1104734}), so the predicted structure is energetically competitive with the reference. In the native order 58 targets are missed at Top-1, 41 of them ranking errors with mostly wrong structure types (the Top-1 candidate has the correct space group in only eight), and the 17 retrieval failures are the same targets.
\FloatBarrier
\subsection{Performance by chemical family}\label{sec:family}
Table~\ref{tab:family_performance} reports the recovery of TCSP3.0 in its default configuration for the families of Supplementary Table~\siref{tab:composition}. On T800, recovery varies widely with chemistry. Compounds containing a rare earth reach 81.7\% at Top-1 and 90.9\% at Top-20, and intermetallics reach 80.7\% and 83.1\%, against 64.8\% and 74.0\% for all 800 targets. Oxygen-containing compounds, which make up 51.0\% of T800, reach 58.6\% and 70.1\%, and halogen-containing compounds reach 54.8\% and 63.0\%. Compounds with a d-block metal (66.8\% and 75.9\%) and with S, Se or Te (66.1\% and 74.0\%) are close to the cohort average. Comparing each family with the remaining targets (Fisher exact test, Bonferroni-corrected over the 24 comparisons made on T180 and T800), the rare-earth contrast is significant at Top-1 and Top-20, the oxygen and intermetallic contrasts at Top-1 and the halogen contrast at Top-20. Oxygen-containing compounds account for 122 of the 208 T800 targets missed at Top-20 (59\%) and halogen-containing compounds for 54 (26\%), whereas rare-earth compounds and intermetallics account for 22 and 14 (11\% and 7\%); the categories overlap, so these shares do not sum to 100\%.

\begin{table}[!htbp]\centering\small\setlength{\tabcolsep}{4.5pt}
\caption{TCSP3.0 recovery by chemical family (default configuration, ORB-v3 relaxation and energy ranking). Entries are the percentage of the family's targets recovered within the Top-1 and Top-20 candidates; $n$ is the number of targets in the family. Families are those of Supplementary Table~\siref{tab:composition} and overlap, so rows are not additive.}
\begin{tabular}{lrrrrrrrrr}\toprule
 & \multicolumn{3}{c}{T46} & \multicolumn{3}{c}{T180} & \multicolumn{3}{c}{T800}\\\cmidrule(lr){2-4}\cmidrule(lr){5-7}\cmidrule(lr){8-10}
Chemical family & $n$ & Top-1 & Top-20 & $n$ & Top-1 & Top-20 & $n$ & Top-1 & Top-20\\\midrule
Contains oxygen & 16 & 75.0 & 81.2 & 40 & 62.5 & 87.5 & 408 & 58.6 & 70.1 \\
Intermetallic & 14 & 85.7 & 100.0 & 46 & 78.3 & 91.3 & 83 & 80.7 & 83.1 \\
Contains a rare earth & 15 & 86.7 & 100.0 & 52 & 78.8 & 96.2 & 241 & 81.7 & 90.9 \\
Contains a d-block metal & 36 & 77.8 & 88.9 & 137 & 76.6 & 92.0 & 452 & 66.8 & 75.9 \\
Contains a halogen & 6 & 66.7 & 100.0 & 26 & 57.7 & 84.6 & 146 & 54.8 & 63.0 \\
Contains S, Se or Te & 8 & 75.0 & 87.5 & 27 & 59.3 & 85.2 & 177 & 66.1 & 74.0 \\
\midrule
All targets & 46 & 76.1 & 89.1 & 180 & 72.2 & 90.6 & 800 & 64.8 & 74.0 \\
\bottomrule\end{tabular}\label{tab:family_performance}\end{table}

On T180 the families are smaller (26 to 137 targets) and the ordering is similar, but only the d-block contrast at Top-1 (76.6\% against 58.1\%) reaches nominal significance, and it does not survive the correction. The lower T800 rates are only partly a matter of family mix: oxygen-containing compounds are 51.0\% of T800 but 22.2\% of T180, yet within families T800 is also lower, for example at Top-20 for oxygen-containing (70.1\% against 87.5\%) and for all other compounds (78.1\% against 91.4\%). A plausible explanation, which we did not test, is that rare-earth compounds and intermetallics form isostructural series that the template library covers densely, so that substitution from a close analogue succeeds, whereas oxides and halogen compounds more often adopt structures for which no close analogue is available or for which the oxidation-state assignment needed for the substitution is less certain. T46 has only 6 to 37 targets per family and is reported for completeness.

\FloatBarrier
\section{Discussion}
TCSP3.0 combines complementary template retrieval with chemically constrained element mapping to predict crystal structures from composition. It transfers known crystal frameworks to other chemistries through coefficient-preserving mappings, simultaneous species substitution and validation of the output composition and cell. Reusing existing template libraries, element embeddings and oxidation-state models enables prediction without training a model. In its default configuration, with ORB-v3 relaxation and ORB-v3 energy-per-atom ranking of the exported candidates, TCSP3.0 matches the designated reference structure at Top-1 for 72.2\% of T180 and 64.8\% of T800 targets, and within the top 20 candidates for 90.6\% and 74.0\%. In the native chemical order these Top-1 rates are 67.8\% and 64.0\% and the T180 Top-20 match rate is the same 163/180, with the same 17 unmatched targets (Supplementary Table~\siref{tab:orb_current}). These results, together with the improvement in Top-5 structure match rate from structural deduplication, show the value of reliable chemical transfer from reusable structural resources.

The ablations clarify which components support this performance. Embedding retrieval supplies coverage, whereas the PN route changes candidate lists without adding observed Top-20 match rate over embedding-only retrieval on either cohort. Alternative mappings, soft oxidation evidence and radius scoring also change candidates and their orderings, with small or null effects on the match rate under the tested settings. The diagnosis of the original implementation and the proximity of the repaired internal baseline to the full configuration indicate that correctness safeguards account for much of the improvement over TCSP2.0. Structural deduplication provides the clearest additional benefit by improving Top-5 structure match rate on both T180 and T800.

Candidate selection depends on the intended outcome. Deduplication improves the early structure match rate, but tolerant geometric matching can merge candidates with different SG labels; removing deduplication consequently improves Top-20 same-candidate structure-and-symmetry match rate on T800. The refinement comparisons likewise separate geometry changes from ranking changes. MatterSim energy ranking increases the observed Top-1 match rate while reducing the Top-20 match rate, whereas fixed-order relaxation slightly reduces the structure match rate. In the default configuration, ORB-v3 relaxation and energy ranking increase the observed Top-1 and Top-5 match rates on all three cohorts while leaving the Top-20 match rate within one target of the native order. These results motivate selection rules that explicitly account for the desired balance between geometric match rate, symmetry agreement and energy ordering.

There are several limitations of our algorithm and the benchmark study.
The present benchmarks provide a foundation for evaluating TCSP3.0 across broader chemical spaces and expanded structural libraries. However,
TCSP3.0 currently matches only one reference structure per target and does not predict all reasonable polymorphs, novel prototypes or physical stability. Force convergence and low learned-potential energy likewise require separate interpretation and training-overlap assessment~\cite{CHGNet2023,M3GNet2022}. Another limitation is that our evaluation of PNcsp+ uses only our adaptation, which draws on a different template inventory without formation-energy filtering and ranks unrelaxed candidates by single-point energies (as the original PNcsp+ does), whereas TCSP3.0 and the deep generative models share the ORB-v3 relaxation and energy ranking~\cite{PNcsp2026,Shotgun2024}. Also, the algorithms' runtimes were measured on a shared cluster with heterogeneous nodes and include model loading, but they show the general trend.

\paragraph{Future directions}
Our algorithm can be further improved by achieving better chemical context and candidate selection. Site-resolved mixed valence and bond-valence validation could improve the assessment of local chemical environments, while calibrated uncertainty and topology clustering of the template library could guide candidate allocation and selection. Fixed-library paired comparisons can test these extensions by measuring changes in coverage and ranking separately. In addition, additional performance metrics can be introduced to consider geometry, volume, symmetry, energy and stability measurements to complement ranked reference matching.
For discovery applications, predicted structures can serve as starting structures for MLIP or first-principles refinement followed by thermodynamic and dynamical assessment. It is also interesting to combine template-based CSP with coordinate-generating models, which offers a way to explore compositions for which suitable structural analogues are scarce.
Overall, reliable template transfer provides a starting point for composition-driven structure searches. Combining curated structural resources with better candidate selection and independent physical evaluation offers a path from reference matching toward broader materials-discovery applications.

\section{Conclusion}
TCSP3.0 corrects the element-mapping and substitution errors that gave one in five TCSP2.0 predictions the wrong composition, and adds embedding and PN retrieval, multiple mappings, soft oxidation evidence, radius compatibility and structural deduplication. With ORB-v3 relaxation and energy ranking, TCSP3.0 achieves Top-1/5/20 match rates of 76.1/89.1/89.1\% on T46, 72.2/88.3/90.6\% on T180, and 64.8/72.4/74.0\% on T800. Without relaxation, the corresponding match rates in native chemical order are 73.9/84.8/89.1\%, 67.8/85.6/90.6\%, and 64.0/71.9/73.9\%, respectively. This doubles TCSP2.0's T180 Top-20 match rate and exceeds PNcsp+, CrystalFlow and MaskGXT at every cutoff, and even without relaxation the native TCSP3.0 matches 122 T180 targets at Top-1 and 163 at Top-20 in a median of 28~s per target on two CPU cores. Ablations attribute most of the gain to the correctness fixes, with structural deduplication improving early ranking and embedding retrieval providing coverage.

\section*{Data and code availability}
Benchmark definitions, reference and input manifests, prediction lists, source inventories, overlap analyses, experiment protocols and evaluation scripts are retained in the accompanying reproducibility records. Software source is available at \url{https://github.com/usccolumbia/TCSPV3}. The reported native feature outcomes use the frozen V3 implementation and protocol. The default configuration (ORB-v3 relaxation and ranking) is the default of the public package from release 3.1.0; \texttt{-{}-ranker native} returns the native order. Template and pretrained-model assets are subject to their upstream distribution terms.

\section{Contribution}
Conceptualization, J.H.; methodology, Y.F., J.H., L.W., Je.H., L.T.; software, Y.F., J.H.; resources, J.H.; writing--original draft preparation, Y.F., L.W., J.H., M.M., L.T., D.G., H.E.; writing--review and editing, Y.F., J.H., L.T., M.M., D.G., H.E.; visualization, Y.F.; supervision, J.H.; funding acquisition, J.H. Here J.H. denotes Jianjun Hu and Je.H. denotes Jeffrey Hu.

\section{Declaration of conflict of interests}
None

\section*{Acknowledgement}
The research reported in this work was partially supported by the National Science Foundation under grants 2311202 and 2320292. The views, perspectives and content do not necessarily represent the official views of the NSF. The authors gratefully acknowledge the computational resources provided by the Theia high performance computing cluster at the University of South Carolina which is supported by National Science Foundation Major Research Instrumentation Grant No. 2320292. We also acknowledge the technical assistance and resources provided by Research Computing at the University of South Carolina (RRID:SCR\_027488).

\clearpage
\appendix
\section{Substitution-method comparison}
Table~\ref{tab:lineage} summarizes how the substitution-based methods discussed in the Introduction differ in retrieval, mapping and ranking.

\begin{table}[H]\centering\small
\caption{Element-substitution lineage and the stage emphasized by each method. Physical assessment and reported budgets depend on the original publication protocol; these descriptions are not a controlled accuracy comparison.}
\begin{tabularx}{\linewidth}{l>{\raggedright\arraybackslash}X>{\raggedright\arraybackslash}X>{\raggedright\arraybackslash}X}\toprule
Method & Retrieval or similarity & Mapping or ranking & Physical budget / comparison caution\\\midrule
Hautier / pymatgen & Data-mined ionic pair probabilities; supplied templates & Conditional-product threshold, charge checks, optional duplicate filtering & Substitution routines do not themselves relax candidates\\
CSPML & Learned composition-pair classification selects templates & Coefficient-preserving and descriptor-resolved transfer; DFT refinement & Pair-classification accuracy differs from the CSP match rate\\
ShotgunCSP & Template generator GT (with template-free generator GW) & Equal-coefficient substitution and descriptor assignment; learned CGCNN screening and physical assessment of selected candidates & GT/GW generation, target-specific DFT learning and refinement; task-specific budget\\
TCSP1/2 & Composition templates, inherited embedding/chemistry & Oxidation evidence and chemistry; optional physical stages & Prior rates use their own references, exclusions and relaxation\\
PNcsp+ & PN neighbors in OQMD & Two-potential single-point mean; separate SG best list & Unrelaxed benchmark; different library and candidate counts\\
TCSP3.0 & Embedding and PN routes & Multiple mappings, soft neutral evidence, radius score, structural diversity & Pool cap 100; default ORB-v3 relaxation and energy ranking of the top 20 (native order: no relaxation)\\\bottomrule
\end{tabularx}\label{tab:lineage}\end{table}

\clearpage
\bibliographystyle{unsrtnat}
\bibliography{references,references_additional}
\end{document}